\documentclass[letterpaper, 10 pt, conference]{ieeeconf}  % Comment this line out if you need a4paper

\IEEEoverridecommandlockouts                              % This command is only needed if 
\usepackage{amsmath} % assumes amsmath package installed
\usepackage{amssymb}  % assumes amsmath package installed
\usepackage{blindtext}
\usepackage[pdftex]{graphicx}
\usepackage[pdftex]{xcolor}
\usepackage{subcaption}
\usepackage{tikz}
\usepackage{units}

\usepackage{epstopdf}

\usepackage{bm}
\DeclareMathAlphabet{\mathsfit}{\encodingdefault}{\sfdefault}{m}{sl}
\SetMathAlphabet{\mathsfit}{bold}{\encodingdefault}{\sfdefault}{bx}{sl}
\newcommand{\randVec}[1]{\bm{\mathsfit{#1}}}

\title{\LARGE \bf
An Unsupervised Random Forest Clustering Technique\\for Automatic Traffic Scenario Categorization
}

\author{Friedrich Kruber$^{1}$, Jonas Wurst$^{1}$ and Michael Botsch$^{1}$% <-this % stops a space
\thanks{$^{1}$Technische Hochschule Ingolstadt, Research Center CARISSMA, Esplanade 10, 85049 Ingolstadt, Germany,
\{firstname.lastname\}@thi.de}}

\begin{document}

\maketitle
\thispagestyle{empty}
\pagestyle{empty}

%%%%%%%%%%%%%%%%%%%%%%%%%%%%%%%%%%%%%%%%%%%%%%%%%%%%%%%%%%%%%%%%%%%%%%%%%%%%%%%%
\begin{abstract}
A modification of the Random Forest algorithm for the categorization of traffic situations is introduced in this paper. The procedure yields an unsupervised machine learning method. The algorithm generates a proximity matrix which contains a similarity measure. This matrix is then reordered with hierarchical clustering to achieve a graphically interpretable representation. It is shown how the resulting proximity matrix can be visually interpreted and how the variation of the methods' metaparameter reveals different insights into the data. The proposed method is able to cluster data from any data source. To demonstrate the methods' potential, multiple features derived from a traffic simulation are used in this paper. 

The knowledge of traffic scenario clusters is crucial to accelerate the validation process. The clue of the method is that scenario templates can be generated automatically from actual traffic situations. These templates can be employed in all stages of the development process. The results prove that the procedure is well suited for an automatic categorization of traffic scenarios. Diverse other applications can benefit from this work.

\end{abstract}

%%%%%%%%%%%%%%%%%%%%%%%%%%%%%%%%%%%%%%%%%%%%%%%%%%%%%%%%%%%%%%%%%%%%%%%%%%%%%%%%
\section{INTRODUCTION}
With the increasing automation of vehicles the demand for mileage driven during the validation phase increases dramatically. Major efforts are being made to increase the simulation part. On the other hand, field tests are highly valuable in finding deficiencies which are not defined within the specifications \cite{FOTNetData.2016}. They form an integral part of the release process. Car manufacturers need to find new approaches to overcome the ``Approval-Trap'' for automated driving functions \cite{Wachenfeld.2016}. One fundamental step is to memorize and structure the experienced and representative driving situations. Therefore, the vast amount of sensor data needs to be shrinked. This can be achieved by representing a traffic situation with a set of relevant features. These features can then be used in machine learning algorithms for analysis purposes.
 
The machine learning training process of recorded traffic situations is usually manually labeled in order to run supervised classification algorithms. This paper introduces an unsupervised learning procedure for the categorization of traffic situations and demonstrates its potential. The data set in this paper is focused on critical scenarios. However the method can be utilized to cluster all types of traffic scenarios. The method offers an intuitive macro- or microscopic view on the data set by varying only one single metaparameter. 
The proposed clustering method is able to group traffic scenarios from any kind of data source. The input data can be obtained from vehicle sensors, as well as from other sources, like navigation data or information from vehicle2X-communication.
The method can also be applied in other research topics like crash severity predictions \cite{Mueller.2016}. Further thoughts on other applications will follow in Section \ref{conclusions_section}. 

Two related research projects will be mentioned. In \cite{DBLP:journals/corr/ZhaoP17aa} the authors describe an approach to accelerate automated vehicle testing. The authors emphasize the necessity to reduce the testing effort by selecting rare traffic situations, which they specify as critical scenarios. A six-step approach is introduced. Basically, a large amount of real-world data is reduced, so that meaningful scenarios remain. Next, human driver behavior is modelled and used as random variables. The scenarios are then fed into a Monte Carlo test to enlarge the number of scenarios. Finally, a reverse statistical analysis is conducted to check how the automated vehicle would perform, based on real world occurrences of critical scenarios. Another research project spans an overall scope on the testing of highly automated driving functions \cite{ForschungsprojektPEGASUS.2017}. The project aims a widely accepted and standardized validation framework. As in the aforementioned project, relevant traffic scenarios are considered to be the basis for validating automated driving functions.  Moreover, also in \cite{PreventiveandActiveSafetyApplications.2009}  it is recommended to identify driving situations for the validation strategy.
 
This paper demonstrates a machine learning based method which aids in finding the relevant scenarios. The following section outlines the overall architecture of the proposed method. Afterwards the data and feature generation is introduced. Section IV presents the proposed clustering technique based on the Random Forest (RF). The related work of the clustering method, as well as the required background, is introduced within Section IV. The paper is then continued by discussing the clustering results applied to the data set of traffic scenarios. Finally, a conclusion of the presented work is drawn.
\section{ARCHITECTURE}
\begin{figure}
\par\medskip
	\begin{footnotesize}
		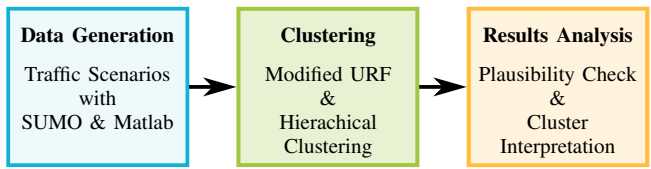
	\end{footnotesize}
	\caption{The overall architecture}
	 \label{fig:architecture}
	 \vspace{-0.5cm}
\end{figure}

This section provides an overview of the architecture as depicted in Fig. \ref{fig:architecture}. The data set of critical scenarios is generated by the usage of the traffic simulation suite SUMO \cite{SUMO_URL}. The data is then processed in Matlab and feature vectors are built (Section \ref{data_section}). 
A description of a traffic situation includes the environmental perception. Modern cars provide various features as GPS data, road details, object lists, rain detection and ambient temperature. These are a few examples of features that can be used for the proposed method. To demonstrate the potential of the proposed unsupervised machine learning procedure, two types of features are chosen. One type contains object behavior features, the second type road infrastructure features. Here, objects are represented by attributes, i.\,e. state-variables, of the simulated vehicles. Road characteristic features are extracted from the roadmap file. For this paper the number of features is limited to ten. Generally, the choice of features should be adjusted to the object under test.
The clustering method is independent of the data source. The features can be constructed by raw sensor data or processed high level data, for example a list of objects surrounding the ego vehicle.

The feature vectors are fed into the clustering process. This work proposes a modified Unsupervised Random Forest (URF) to achieve a data adaptive similarity measure. Some modifications of the RF algorithm enable its use for unsupervised learning in high dimensional spaces. A traffic scenario is represented by a point in a high dimensional feature space. The similarity measure yields a normalized similarity measure between all datapoints. Structuring the relationships is achieved using hierarchical clustering. An optimization step, namely the Optimal Leaf Order (OLO) algorithm, can be performed afterwards. For analyzing the inherent data structure this paper suggests a graphical representation of the data. The graphic implies two advantages. First, it literally shows the clusters. Second, it highlights the similarity measure using a colormap, even for the inter-relationships of separate clusters. Further details regarding the clustering process are described in Section \ref{clustering_section}.

\section{DATA GENERATION}
\label{data_section}
This section describes the procedure of data generation to train the modified URF. It is divided into three parts: the traffic generation and a description of the roadmap including the sceneries. The final part deals with the feature generation. 

\subsection*{Traffic simulation}
In order to generate traffic flow data the simulation suite SUMO is used. SUMO is capable of modelling the behavior of intermodal traffic participants on a microscopic level. The basic requirements to run a simulation imply a roadmap and defined routes for the vehicles. To generate routes, first the characteristics of the vehicles are set up via vehicle types. A vehicle type is defined by several parameters like the maximum acceleration, braking abilities and the minimum reaction time.  Next, a journey for each individual vehicle is generated by defining a string of location points in the roadmap. The final route for an individual vehicle consists of the location points and a randomly associated vehicle type.

The traffic density is varied in order to obtain a medium to high traffic flow up to traffic jams. For the hourly flow rate the guidelines from \cite{ForschungsgesellschaftfurStraenundVerkehrswesen.2015} are consulted. The short-term flow rate, in other words the number of vehicles in the bounded region of interest, is randomized.  

Generally the driver models in SUMO aim towards an accident free traffic flow. Driver models, which adapt their behaviour to other participants are described in \cite{Vanderhulst.1999}. If the modelled leader vehicle determines the behaviour of the following vehicles, the driver model is named a car-following model, which is the basis of SUMOs vehicle behaviour. The vehicles in the simulation keep safe distances and are center-orientied within the lane width. To generate critical scenarios and to overcome the save driving behaviour, several parameters like the reaction time are tweaked. Further details regarding adjustable parameters are listed on the developers support page. Breaking traffic rules at junctions is another effective way to increase the probability of critical scenarios. This approach is supplemantary used by adjusting the rules locally within the roadmap file. According to real traffic scenarios, the violation of rules like disregarding the right of way or undercutting save gaps cause accidents. In addition, the sublane model introduced by \cite{Semrau.2016} is used. The model enables an aggressive longitudinal approach and critical lane change behaviours. The car-following model is then switched to a modified version of \cite{Krau.1998}. The integrated lane change model considers the dependencies of longitudinal and lateral movements. While running the simulation, an online interaction can be established with the interface developed by \cite{Acosta2015}. The interface is used to gather vehicle data and to set commands in order to influence the vehicle's behavior.

\subsection*{Roadmap and sceneries}
For this paper some parts of the road network of the city of Ingolstadt and the German highway A9 are created and modified using the SUMO importer ``netconvert''. The raw file of the roadmap is provided by openstreetmaps. 
Eight sceneries within the extracted map are chosen for this work. A scenery is here defined by the road characteristics. The analyzed sceneries imply merging situations on the German Highway, inner-city crossings and roundabouts with a varying number of lanes. To speed up the simulation time, only data from the depicted sceneries are logged. For that reason, the point of interest is defined with a radius in front and behind the junction point. All the vehicles inside that bounded region are tracked for the data generation process. The radius differs from \unit[15]{m} to \unit[50]{m}, depending on the expected average velocities. The next process step handles a threshold algorithm to seperate the critical scenarios from the non-critical ones. 

\subsection*{Data selection and feature generation}
The analyzed data in this work is restricted to critical scenarios. Here, the definition of criticality applies to scenarios, where the vehicles reach a close relative distance, not necessarily leading to a collision. In that event the interaction between two vehicles is taken into account, even though scenarios with several participants occur. The binary criticality index $k \in [0,1]$ is introduced. 

The timespan for a scenario is chosen to be $\Delta\,t= t_{0} - t_{-2} = \unit[2.0]{s}$. The scenario starts at the time step $t_{-2}$. From that moment the most likely trajectory hypothesis is used for $t_{0}$. This work does not focus on trajectory planning. Interested readers are referred to the survey \cite{Lefevre.2014}. Here, the most likely hypothesis is the trajectory calculated by SUMO itself. This ensures, that all objects stay within the road boundaries. Apart from the changes described above, the vehicles adapt to other participants to avoid collisions. On the downside the accuracy of the trajectory planning is limited due to the simple structure of the kinematic model. The obtained vehicle data is a set of state variables (position, velocity and orientation) plus the vehicle boundaries. 

As millions of scenarios shall be simulated, a filter which allows fast selection is implemented.
The pre-filter eliminates the uncritical scenarios with a rough thresholding using the relative distance and velocities of two objects.
Afterwards the relevant events are captured in the main routine by using two criteria. To pass the first criterion, a physical contact within the next \unit[0.3]{s} after $t_{0}$ will most likely occur. The calculation is based on a kinematic model with constant acceleration and orientation vector for both vehicles within the next \unit[0.3]{s} ($t_{0.3}$). The objects are modeled as polygons. The algorithm checks for an overlap between the polygons. The assumption allows fast computing and supposes that in most cases drivers do not change the dynamic properties significally within the last \unit[0.3]{s} before a collision occurs \cite{Zoller.1998}. If a collision will not occur, the second criterion checks for a close distance $d_\mathrm{rel}$ and a remaining velocity $v$ for both vehicles: 
\begin{equation}
k(t_{0.3})=\left\lbrace \begin{array}{c l}
1 & \lbrace TTC = \unit[0]{s} \rbrace \\
1 & \lbrace d_\mathrm{rel} < \unit[0.3]{m} \wedge v > \unit[2]{m/s} \rbrace \\
0 & \lbrace \mathrm{otherwise} \rbrace
\end{array}\right.
\end{equation}
Given $k(t_{0.3}) = 1$ the scenario is chosen to be relevant. By tightening the probability of generating critical scenarios, in total 22\,519 scenarios are selected within 2.2 million simulated seconds.
	
The applied features are separated into two layers. The road-layer contains three features, which can be obtained by modern vehicles: the number of lanes ($n_{L}$), the speed limit ($v_\mathrm{lim}$) and the approximated radius of the road segment ($r$) at $t_{0}$. Radius values above \unit[7000]{m} are rarely built according to \cite{ForschungsgesellschaftfurStraenundVerkehrswesen.2015} and set to $r = \unit[11111]{m}$ as a replacement for straight roads. 
The object layer contains in total seven features, of which four present the velocities $v$ of both vehicles at $t_{0}$ and $t_{-2}$. They comprise the absolute, the relative and the change of the velocities. The fifth object-feature is given by the relative angle between the vehicles ($\delta_\mathrm{rel}$) at $t_{0}$ and varies from \unit[0]{$^{\circ}$} to \unit[180]{$^{\circ}$}. That feature allows conclusions for the collision constellation as rear-end or side crash situations. The last two features indicate if one or both vehicles performed a braking maneuver during the scenario ($b$). They pronounce the change of velocities on a binary level, where 1 equals a decrease in velocity between $t_{-2}$ and $t_{0}$. This enhances the clustering and helps to analyze the clusters as described later on in Section \ref{results_section}. Additionally, a braking maneuver indicates the awareness of the driver. Environmental features are not supported by the simulation tool. The awareness of the driver and the aspect of the road friction coefficient can only be parameterized on a high level by setting the reaction time and the braking abilities of a vehicle. These parameters are set in a random order in the vehicle types described above. The two vehicles are addressed by eg for the ego and tg for the target vehicle.
In total ten features are selected: 
\begin{equation}
\begin{aligned}
\randVec{x} = [v_{\mathrm{eg}_\mathrm{t-2}},v_{\mathrm{eg}_\mathrm{t0}},b_\mathrm{eg},v_{\mathrm{tg}_\mathrm{t-2}},v_{\mathrm{tg}_\mathrm{t0}},b_\mathrm{tg},\delta_\mathrm{rel},r,v_\mathrm{lim},n_\mathrm{L}]^\intercal.
\end{aligned}
\end{equation}
The three scenery types (highway, crossings and roundabouts) should be well separable with the chosen road features. So the big sized clusters should result mainly from the road features. Smaller clusters should result from the object features as their values vary more. The rather small set of ten features supports an efficient analyzing process to check the potential of the clustering method.
However it should be mentioned, that the proposed method is not limited to a certain number of features.      

\section{CLUSTERING}
\label{clustering_section}
\begin{figure*}
\par\medskip
	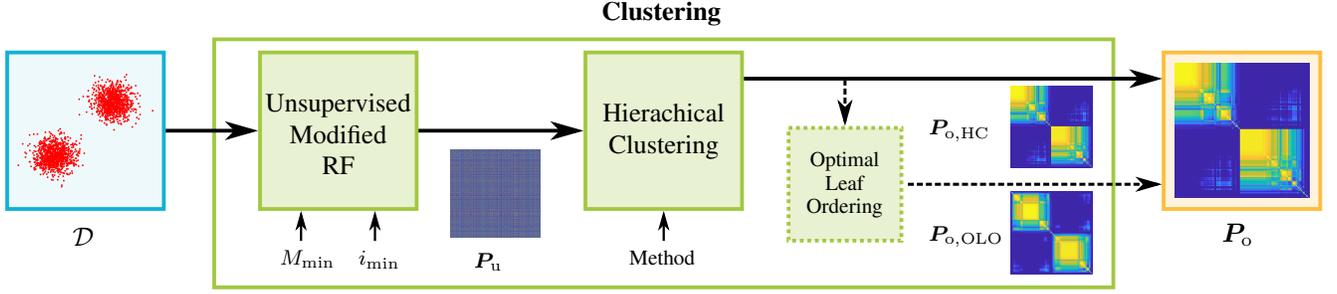
	\caption{Architecture of the introduced clustering technique. Note: The OLO step is optional.}
	\label{fig.ClusteringToolChain}
	\vspace{-0.5cm}
\end{figure*}

This section mainly addresses the uncovering of data inherent structures, e.\,g. of traffic scenarios. Therefore, a modified URF is used to generate a proximity matrix ($\bm{P}_{\mathrm{u}}$ in Fig. \ref{fig.ClusteringToolChain}). By agglomerative hierarchical clustering methods and optionally the OLO algorithm, a reordered proximity matrix ($\bm{P}_{\mathrm{o}}$) is produced. The matrix $\bm{P}_{\mathrm{o}}$ represents the clusters in the data and provides a graphical interpretation of the data inherent structure.

For the mathematical modeling of the clustering problem, it is assumed that the unlabeled set $\mathcal{D}$ is generated by random variables $\randVec{x} \in \mathcal{X}$ ($\mathcal{D}=\left\lbrace \bm{x}_1,\dots,\bm{x}_M \right\rbrace$), where $\mathcal{X}$ is used instead of $\mathbb{R}^Q$, since $\bm{x}$ can also contain categorical variables and $Q$ is the dimension of the vector $\randVec{x}$. The proximity between all datapoints $\bm{x}$ is given by the proximity matrix $\bm{P}$, where $P_{ij}$ is the $ij$th element representing the proximity of datapoints $\bm{x}_i$ and $\bm{x}_j$. Accordingly, $\bm{P}$ is a $M\times M$ symmetric matrix, with $M$ being the number of datapoints. The matrix $\bm{P}$ can be normalized so that $P_{ij} \in [0,1]$. In the following the normalized matrix $\bm{P}$ is considered. A high value of  $P_{ij}$ indicates a high similarity for the datapoints $\bm{x}_i$ and $\bm{x}_j$.

\subsection*{Unsupervised Random Forest and Clustering Tree}
In \cite{Breiman02URF} the usage of the RF\footnote{The RF technique is explained in \cite{Breiman2001random}. Accordingly, it is a randomized ensemble learning method using Classification And Regression Trees (CART) as base learners.} for unlabeled data is introduced. This URF is capable of providing a proximity matrix $\bm{P}$, which can be used for well-known clustering methods\footnote{Since $1-P_{ij}$ can be interpreted as a squared Euclidean distance in a high dimensional space \cite{Breiman02URF}.}. In order to perform unsupervised learning with the RF technique, the well-known procedure that is applied for supervised learning must be changed. Therfore, to turn the unsupervised learning task into a supervised one, two classes, $A$ and $C$ must be defined. All datapoints $\bm{x}$ of $\mathcal{D}$ are labeled with $A$, whereas $C$ is assigned to the points of a synthetic data set $\mathcal{S} = \left\lbrace \bm{z}_1,\dots,\bm{z}_K \right\rbrace ,\bm{z} \in \mathcal{X}$. In \cite{Breiman02URF} the construction of $\mathcal{S}$ is realized by ``independent sampling from the one-dimensional marginal distributions of the original data'' $\mathcal{D}$. Another construction principle is explained in \cite{Shi2006}, where $\mathcal{S}$ is build by sampling randomly from an assumed uniform distribution within the $Q$-dimensional hypercube defined by the minimum and maximum values of $\mathcal{D}$. Independent of the way how the synthetic data is generated, the appearing task is to distinguish between noise, i.\,e. the data set $\mathcal{S}$ and the actual data $\mathcal{D}$. The underlying mechanism is that, if there is a structure in the data in $\mathcal{D}$ the RF has to fit its leaves to it in order to achieve a low error.

One main drawback of the proposed noise generation methods occur with high dimensional input spaces. In order to force the RF to fit properly to the inherent structure, the noise distribution has to be very dense. If there is no more noise data left to perform the dividing task, the resulting leaves become large. Due to the curse of dimensionality \cite{Bishop.2006}, for high dimensional input spaces this leads to a huge amount of necessary synthetic datapoints and accordingly to a highly unbalanced ratio between $A$ and $C$. Even though the sampling from marginal distributions addresses this issue by sampling mainly in the regions of interest, the results in high dimensional spaces might not be satisfactory. In this work a possible solution is presented.

In \cite{Liu2000} a method called CLustering Tree (CLTree) is explained. In this method, the leaves of the decision tree are considered as clusters. However, in \cite{Liu2000} the noise is not explicitly generated, but only the number of points from a uniform distribution is used for the calculations in each split. The number of noise points in a node of the tree is chosen to be equal to the number of datapoints from $\mathcal{D}$ in that node.

Furthermore, in \cite{CGV-035} an unsupervised variant of a RF can be found, called density forest. A modification of the density forest is presented in \cite{Pei2013}.

\subsection*{Modified URF}
In order to achieve viable results with the URF, the idea of generating noise from CLTree is adopted. In the following the resulting modified URF is explained.

\paragraph{Noise Generation}
A RF is considered consisting of $B$ CART trees, where a tree $T_b$ is built based on a bagged data set $\mathcal{D}_b$. Each $T_b$ consists of $N_{b}$ nodes, where a single node $t_{n,b}$ represents a subspace of $\mathcal{X}$, namely $\mathcal{X}_{t_{n,b}} \subseteq\mathcal{X}$. The root node $t_{0,b}$ represents the subspace defined by the respective bagged data set, accordingly $\mathcal{X}_{t_{0,b}}=\mathcal{X}_{\mathcal{D}_{b}}$ where $\mathcal{X}_{\mathcal{D}_{b}}$ represents the hypercube defined by the minimum and maximum values of the bagged data set. The synthetic data set $\mathcal{S}$ is assumed as uniformly distributed at every split and hence for every dimension. It has to be noted that at each split only a subspace with $\tilde{Q}=\sqrt{Q}$ randomly chosen dimensions is used. The required point density of the noise is defined as
\begin{equation}\label{eq.NoiseDensity}
\Delta\mathcal{U}_{t_{n,b,\tilde{q}}} = \frac{M_{\mathcal{D}_b}(\mathcal{X}_{t_{n,b}})}{\mathrm{max}\left\lbrace\mathcal{X}_{t_{n,b}}\right\rbrace_{\tilde{q}}-\mathrm{min}\left\lbrace\mathcal{X}_{t_{n,b}}\right\rbrace_{\tilde{q}}},
\end{equation}
where $\mathrm{max}\left\lbrace\mathcal{X}_{t_{n,b}}\right\rbrace_{\tilde{q}}$ and $\mathrm{min}\left\lbrace\mathcal{X}_{t_{n,b}}\right\rbrace_{\tilde{q}}$ applies the max and min function to the dimension of $\mathcal{X}_{t_{n,b}}$ that is specified by the index $\tilde{q}$, respectively. $M_{\mathcal{D}_b}(\mathcal{X}_{t_{n,b}})$ is the number of datapoints of the bagged data set $\mathcal{D}_b$ belonging to the subspace $\mathcal{X}_{t_{n,b}}$. In order to determine the split which minimizes the Gini impurity \cite{breiman1984classification} the number of noise data has to be calculated for the left ($M_{\mathcal{S},l_{n,b}}$) and the right ($M_{\mathcal{S},r_{n,b}}$) children nodes. It holds that
\begin{align}
M_{\mathcal{S},l_{n,b}}(\tau_{\tilde{q}})&=\Delta\mathcal{U}_{t_{n,b,\tilde{q}}} \left(\tau_{\tilde{q}} -  \mathrm{min}\left\lbrace\mathcal{X}_{t_{n,b}}\right\rbrace_{\tilde{q}}\right)\\
M_{\mathcal{S},r_{n,b}}(\tau_{\tilde{q}})&=M_{\mathcal{D}_b}(\mathcal{X}_{t_{n,b}})-M_{\mathcal{S},l_{n,b}}(\tau_{\tilde{q}}),\label{eq.MSR}
\end{align}
with $\tau_{\tilde{q}}$ being the threshold of the dimension of $\randVec{x}$ specified by $\tilde{q}$. At each split the threshold-dimension combination ($\tau_{\tilde{q}_{\mathrm{opt}},\mathrm{opt}}$) which leads to highest relative Gini impurity reduction is chosen.

\paragraph{Pruning}
The trees of a RF in a classification tasks are normally fully grown. With the balanced noise assumption at each split this can lead to small leaves. In the worst case the leaves can contain only one datapoint from the data set $\mathcal{D}$. Since in this work the RF is used to provide a similarity measure, the trees have to be pruned in order to achieve a valuable proximity matrix $\bm{P}$. In the following two possible pruning techniques are proposed, which are used during the tree growing process.
\begin{itemize}
\item{$M_{\mathrm{min}}$: One possible pruning criteria is the minimum number of real datapoints a node must contain, in order to make a new split. All nodes which consist of less points than the threshold $M_{\mathrm{min}}$ are turned into leaves. A leaf is labeled with the major class ($A$ or $C$).}
\item{$i_{\mathrm{min}}$: Another possible criterion is the minimum Gini impurity which a node must fulfill in order to make a new split. Like above, all nodes which fulfill the condition $i_{n,b}<i_{\mathrm{min}}$ are turned into leaves, where $i_{n,b}$ is the Gini impurity of the $n$th node in the $b$th tree.}
\end{itemize}

%Criticism of pruning criterias

The two termination criteria can be used in a combined manner. Those metaparameters have to be chosen depending on the data set which is investigated.

After training the URF, the proximity matrix $\bm{P}$ has to be determined. For each possible datapoint pair it is counted in how many trees they end up in the same leaf. These counts are normalized by the number of trees ($B$). If the resulting value $P_{ij}$ is equal to one, the $i$th and $j$th datapoints terminate in the same leaf in every tree -- if zero not in a single leaf. So $P_{ij}$ is a data-adaptive similarity measure \cite{Breiman02URF}.
%which for what?

\subsection*{Matrix Reordering}
In this work the clustering is realized by a combination of a proximity matrix reordering task and visual interpretation. The ordered proximity matrix is capable of proving deep insights into the data structure and even into inter-cluster and intra-cluster relationships. If one is interested in cluster identification it is recommended to use hierarchical clustering approaches in order to perform the reordering \cite{Behrisch2016}.

The interpretation of the reordered proximity matrix is done visually, based on the appearance of basic patterns. A detailed overview of typical patterns can be found in the survey \cite{Behrisch2016}. In this work the proximity matrix is shown as a square image, where dark blue pixels represent zero entries ($P_{ij}=0$) and yellow pixels one entries\footnote{For further descriptions see Matlabs parula colormap.}. One can think about bright squares along the diagonal representing clusters.

Agglomerative hierarchical clustering methods are defined for distances between datapoints. Therefore, the proximity matrix $\bm{P}$ has to be transformed into the dissimilarity matrix $\bm{D}$, where $D_{ij}=\sqrt{1-P_{ij}}$ holds.
Explanations about the possible linkage methods as well about the different implementations can be found in the survey \cite{Murtagh1983}.

As shown in Fig. \ref{fig.ClusteringToolChain}, reordering the unsorted proximity matrix $\bm{P}_{\mathrm{u}}$ is performed by a row- and columnwise permutation. Therefore, the datapoint index order determined by the hierarchical clustering is used to generate $\bm{P}_{\mathrm{o,HC}}$.

If the resulting structure in $\bm{P}_{\mathrm{o,HC}}$ is not satisfactory, the datapoint index order can be further refined by the OLO algorithm \cite{bar2001fast}, namely $\bm{P}_{\mathrm{o,OLO}}$. This method aims to minimize the distances ($D_{ij}$) between successive datapoints.

In Fig. \ref{fig.ClusteringToolChain} the whole clustering technique along with the results for an example data set $\mathcal{D}$, where $M=2000$, is shown. The used settings for the shown results are $B=200$, $M_{\mathrm{min}}=2000$ and the average linkage method. Here the criteria $M_{\mathrm{min}}$ and $i_{\mathrm{min}}$ are combined with a logical OR, therefore only $i_{\mathrm{min}}$ is used for pruning. In $\mathcal{D}$ one can identify two clusters. Assuming that the order of datapoints in $\mathcal{D}$ is random, one obtains $\bm{P}_{\mathrm{u}}$ as the output of the modified URF. Reordering $\bm{P}_{\mathrm{u}}$ by average linkage leads to $\bm{P}_{\mathrm{o,HC}}$, which already contains two big bright squares along the diagonal. Those are the corresponding representatives of the two clusters in $\mathcal{D}$. Refining $\bm{P}_{\mathrm{o,HC}}$ by the OLO yields $\bm{P}_{\mathrm{o,OLO}}$. Like before, two big bright squares are visible which are matching to the clusters. Though, the inner order of the squares in $\bm{P}_{\mathrm{o,HC}}$ and $\bm{P}_{\mathrm{o,OLO}}$ has changed, whereby in each square of $\bm{P}_{\mathrm{o,OLO}}$ one big yellow square occurs. Each yellow square corresponds to the respective cluster center.

In this work only the hyperparameter $i_{\mathrm{min}}$ is used, since its data adaptiveness is superior to $M_{\mathrm{min}}$ for the analyzed data sets. As for the RF-algorithm the larger the number of trees $B$ is, the better. However, after a certain number of trees the improvements are negligible.

The presented clustering technique is able to uncover data inherent structures and presents them in an interpretable form.
%image processing to determine clusters atuomatically semi auto (ref)
%time/storage complexity

\section{RESULTS}
\label{results_section}
\begin{figure}
\par\medskip
\centering
\captionsetup[subfigure]{labelformat=empty}
\begin{subfigure}[b]{0.3\columnwidth}
	\centering
	\caption{$i_\mathrm{min}=0.24$}
	\def\svgwidth{0.95\columnwidth}
	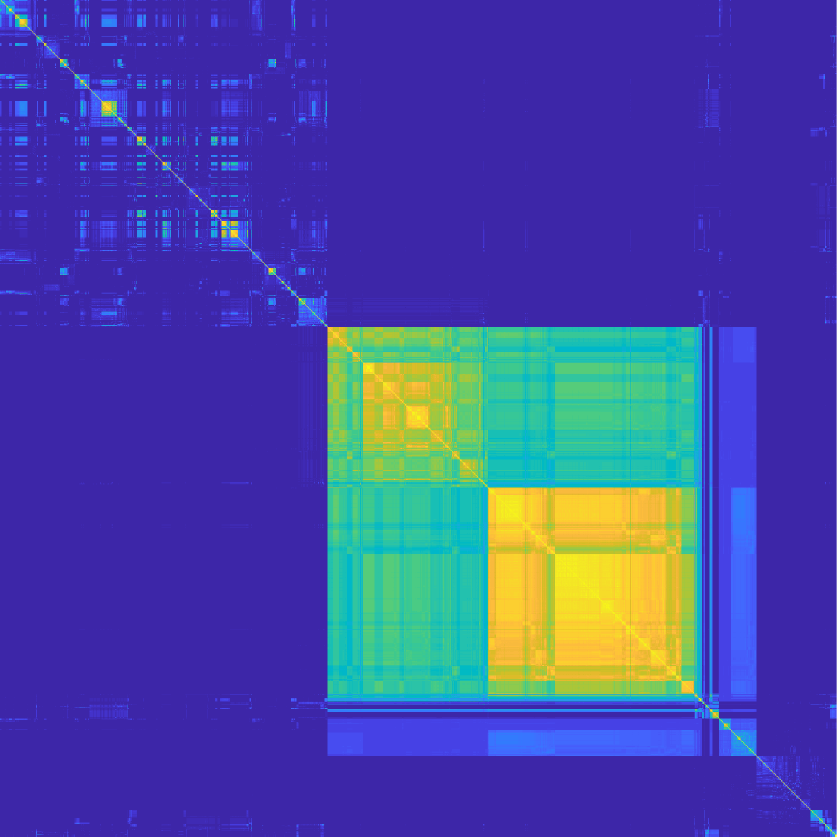
\end{subfigure}
\begin{subfigure}[b]{0.3\columnwidth}
	\centering
	\caption{$i_\mathrm{min}=0.29$}
	\def\svgwidth{0.95\columnwidth}
	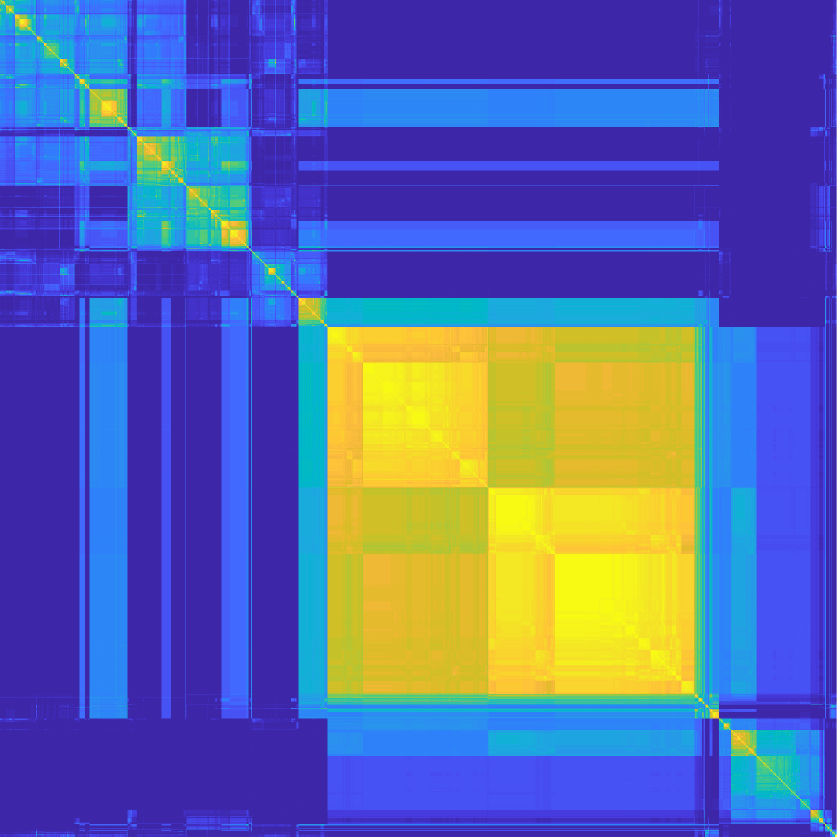
\end{subfigure}
\begin{subfigure}[b]{0.3\columnwidth}
	\centering
	\caption{$i_\mathrm{min}=0.34$}
	\def\svgwidth{0.95\columnwidth}
	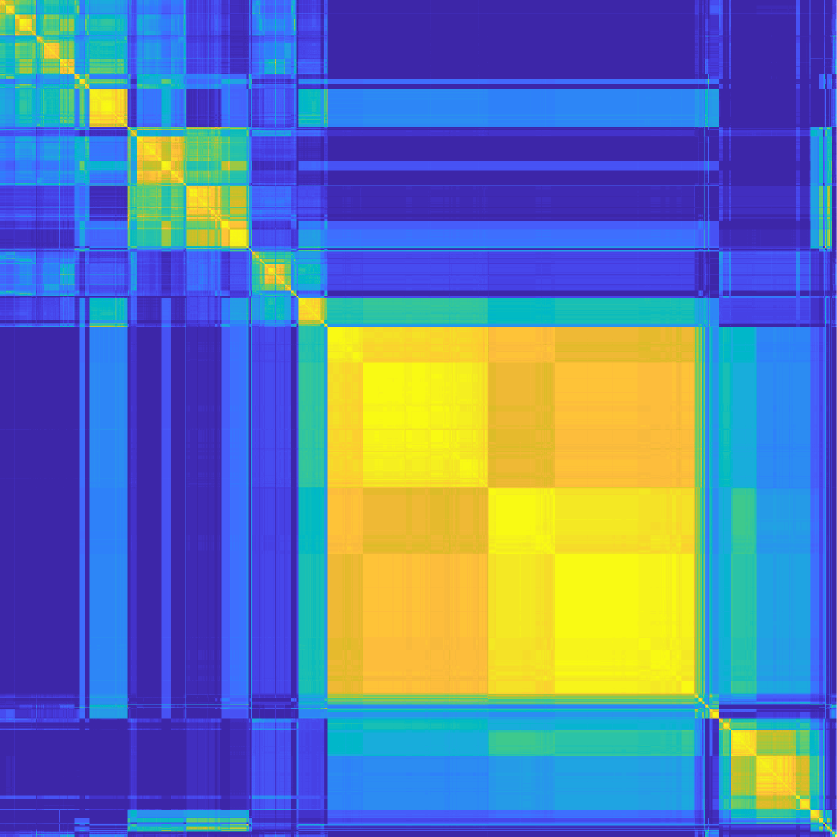
\end{subfigure}
\par\medskip
\begin{subfigure}[b]{0.3\columnwidth}
	\centering
	\def\svgwidth{0.95\columnwidth}
	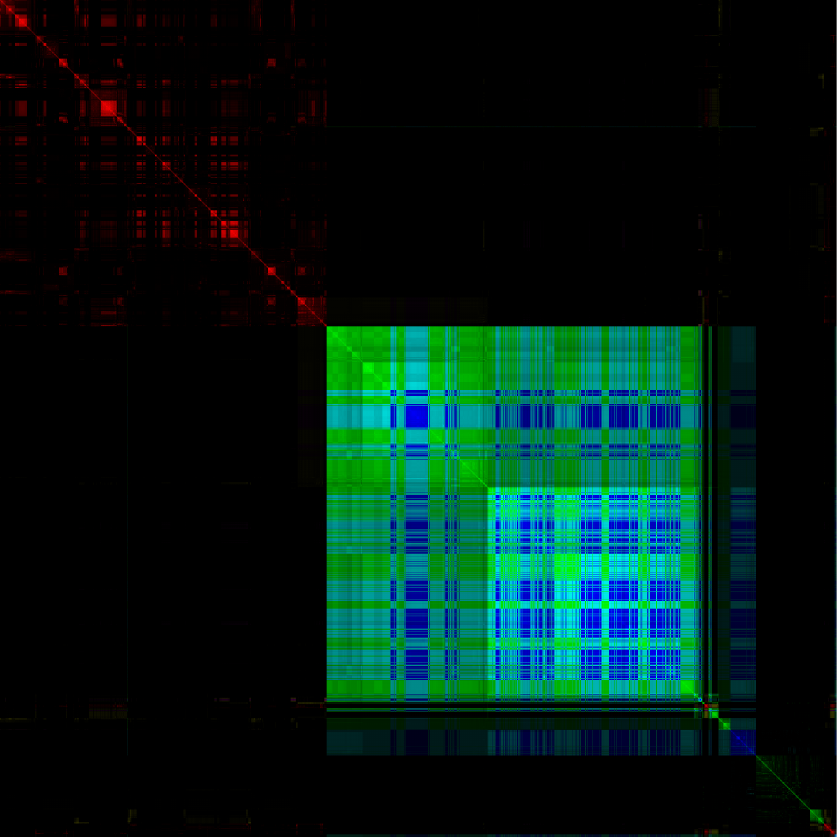
\end{subfigure}
\begin{subfigure}[b]{0.3\columnwidth}
	\centering
	\def\svgwidth{0.95\columnwidth}
	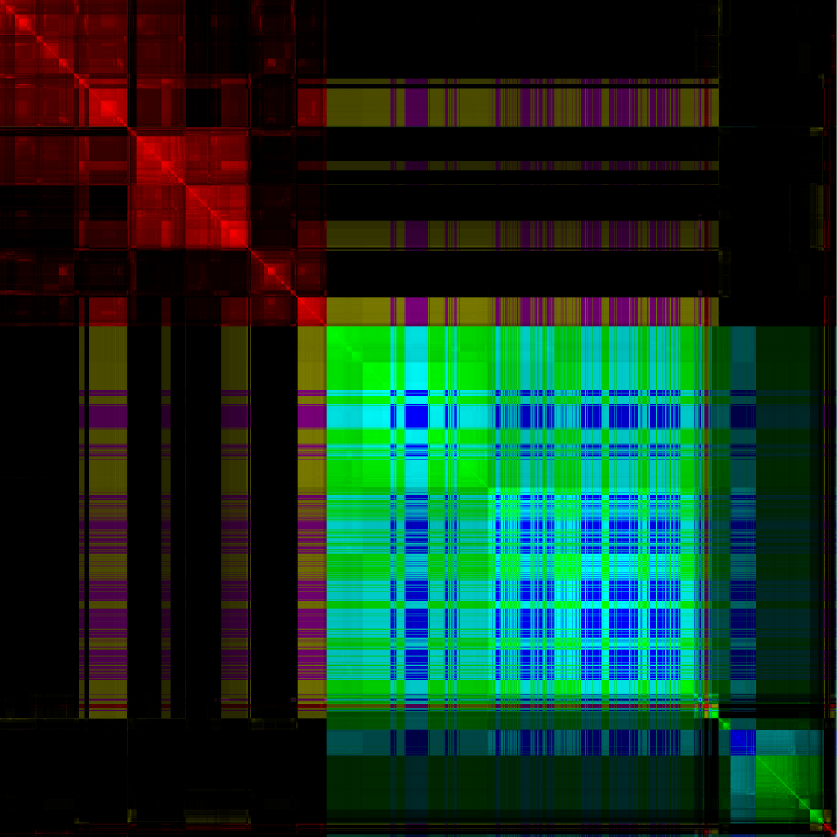
\end{subfigure}
\begin{subfigure}[b]{0.3\columnwidth}
	\centering
	\def\svgwidth{0.95\columnwidth}
	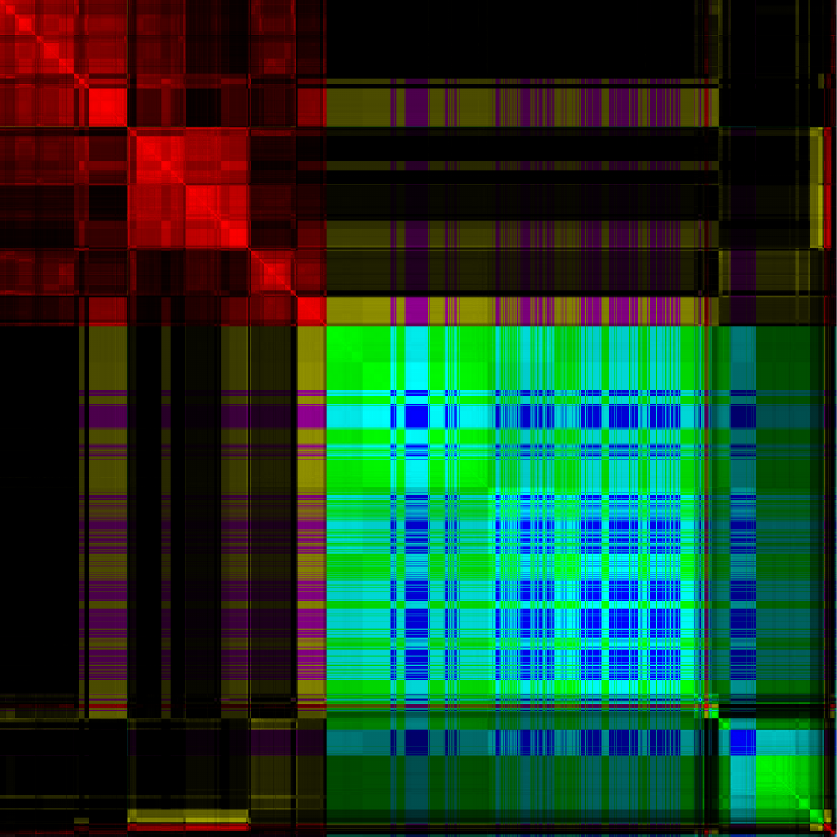
\end{subfigure}
\caption{First row: The overall clustering results varied in three steps of $i_{\mathrm{min}}$. Second Row: To easily evaluate the unsupervised clustering technique, the datapoints are labeled according to their scenario type: highways (red), crossings (green) and roundabouts (blue).}
\label{fig:fig_results}
\vspace{-0.5cm}
\end{figure}

This section focuses on some major findings and is split into two parts: first, results concerning the clustering and second, a discussion on the interpretation on the scenario clusters. For this particular data set no considerable improvements by the OLO algorithm were found. In consequence, the OLO influence is not taken into account for the further discussion.
The clustering parameters are set as follows: the number of trees is $B = 200$, the used linkage method is average (AL) and the metaparameter $i_{\mathrm{min}}$ is varied. Each subfigure in Fig. \ref{fig:fig_results} contains $M\times M = 22\,519 \times 22\,519$ pixels, i.\,e. 22\,519 scenarios.

\subsection*{Clustering results}
The clustering method does indeed reveal similarities by showing squares in different shapes as depicted in Fig. \ref{fig:fig_results}. The varying of the metaparamter $i_{\mathrm{min}}$ changes the structures. The parameter acts like a magnifying glass. A lower value of $i_{\mathrm{min}}$, which is equivalent to a higher threshold regarding the similarity measure darkens the pixels. When pruning the tree at a deeper level, the chances of ending up in different leaves increases and lowers the final similarity measure. Generally higher values of $i_{\mathrm{min}}$ yield larger squares. A suitable fit of the parameter $i_{\mathrm{min}}$ depends on the data set.
The following depicted matrices are square matrices, but due to presentation reasons they are stretched in the horizontal axis.

\subsection*{Scenario categorization results}
The resulting scenario clusters are generated automatically. Nevertheless, they can be interpreted with expert knowledge. The subfigures in the first row in Fig. \ref{fig:fig_results} vary in colors from blue, green, orange to yellow. Blue pixels indicate a low similarity measure between two scenarios. Yellow pixels indicate the contrary, a high similarity value. Green and orange pixels are located in between the normalized scale, respectively. Pixels on the diagonal are yellow colored, because the datapoints are compared to themselves which yields a similarity of one ($P_{ii} = 1$).

To support the discussion, the three subfigures in the second row of Fig. \ref{fig:fig_results} are labeled. Highway mergings are colored red, the crossings green and  blue for the roundabouts. The crossing and roundabout scenarios together will be addressed as inner city scenarios. On a first glance some dominant big squares can be detected. Taking a look at the labeled subfigures, the highway scenarios are clearly separated from to the inner city scenarios for most datapoints. According to the feature description in Section \ref{data_section}, the big sized clusters are mainly determined by the road features.

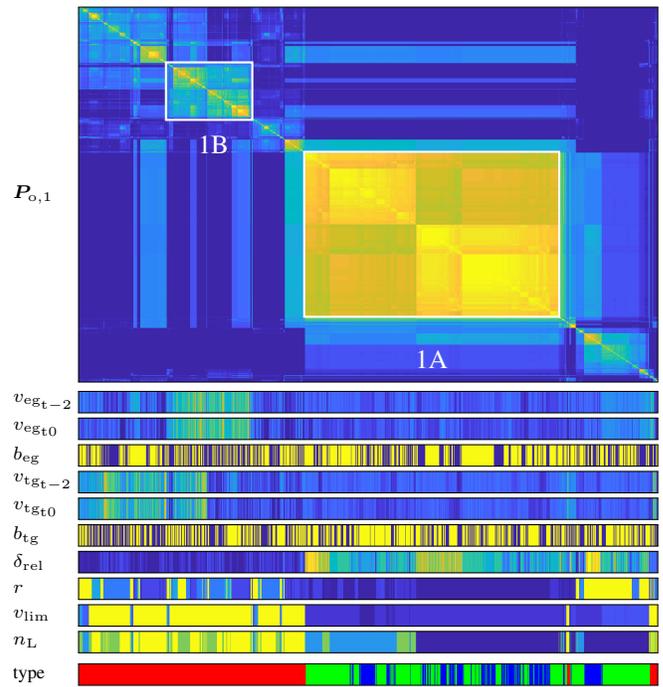
\begin{figure}
\par\medskip
\begin{scriptsize}
	\parbox[c]{0.10\columnwidth}{\raggedright $\bm{P}_\mathrm{o,1}$}%
	\fbox{\parbox[c]{0.89\columnwidth}{\raggedleft \def\svgwidth{0.89\columnwidth} 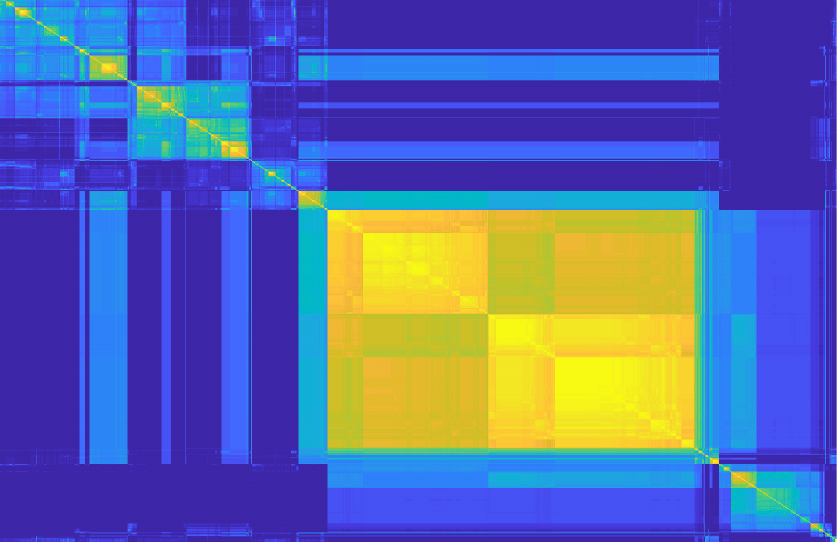}}\\[1ex]
	\parbox[c]{0.10\columnwidth}{\raggedright $v_{\mathrm{eg}_\mathrm{t-2}}$}%
	\fbox{\parbox[c]{0.89\columnwidth}{\raggedleft \def\svgwidth{0.89\columnwidth} 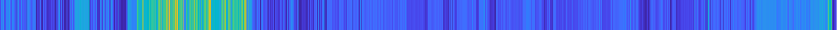}}\\[0.5ex]
	\parbox[c]{0.10\columnwidth}{\raggedright $v_{\mathrm{eg}_\mathrm{t0}}$}%
	\fbox{\parbox[c]{0.89\columnwidth}{\raggedleft \def\svgwidth{0.89\columnwidth} 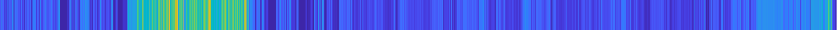}}\\[0.5ex]
	\parbox[c]{0.10\columnwidth}{\raggedright $b_{\mathrm{eg}}$}%
	\fbox{\parbox[c]{0.89\columnwidth}{\raggedleft \def\svgwidth{0.89\columnwidth} 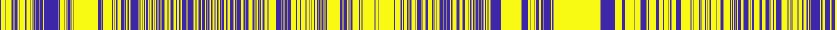}}\\[0.5ex]
	\parbox[c]{0.10\columnwidth}{\raggedright $v_{\mathrm{tg}_\mathrm{t-2}}$}%
	\fbox{\parbox[c]{0.89\columnwidth}{\raggedleft \def\svgwidth{0.89\columnwidth} 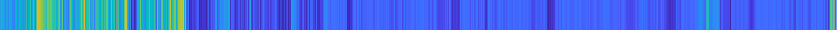}}\\[0.5ex]
	\parbox[c]{0.10\columnwidth}{\raggedright $v_{\mathrm{tg}_\mathrm{t0}}$}%
	\fbox{\parbox[c]{0.89\columnwidth}{\raggedleft \def\svgwidth{0.89\columnwidth} 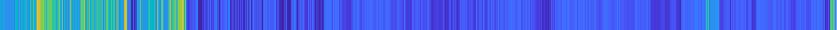}}\\[0.5ex]
	\parbox[c]{0.10\columnwidth}{\raggedright $b_{\mathrm{tg}}$}%
	\fbox{\parbox[c]{0.89\columnwidth}{\raggedleft \def\svgwidth{0.89\columnwidth} 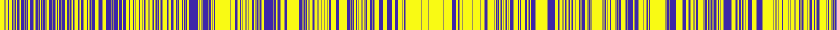}}\\[0.5ex]
	\parbox[c]{0.10\columnwidth}{\raggedright $\delta_\mathrm{rel}$}%
	\fbox{\parbox[c]{0.89\columnwidth}{\raggedleft \def\svgwidth{0.89\columnwidth} 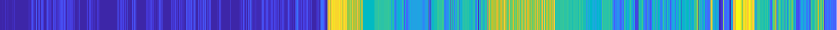}}\\[0.5ex]
	\parbox[c]{0.10\columnwidth}{\raggedright $r$}%
	\fbox{\parbox[c]{0.89\columnwidth}{\raggedleft \def\svgwidth{0.89\columnwidth} 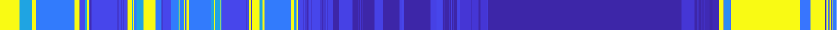}}\\[0.5ex]
	\parbox[c]{0.10\columnwidth}{\raggedright $v_\mathrm{lim}$}%
	\fbox{\parbox[c]{0.89\columnwidth}{\raggedleft \def\svgwidth{0.89\columnwidth} 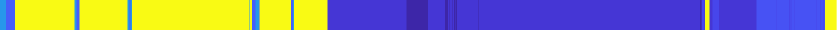}}\\[0.5ex]
	\parbox[c]{0.10\columnwidth}{\raggedright $n_\mathrm{L}$}%
	\fbox{\parbox[c]{0.89\columnwidth}{\raggedleft \def\svgwidth{0.89\columnwidth} 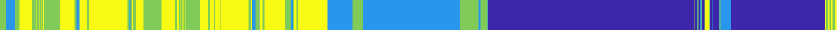}}\\[1.2ex]
	\parbox[c]{0.10\columnwidth}{\raggedright type}%
	\fbox{\parbox[c]{0.89\columnwidth}{\raggedleft \def\svgwidth{0.89\columnwidth} 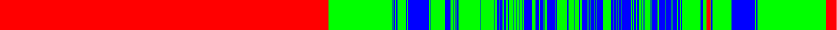}}\\[0.5ex]
	
\end{scriptsize}
\noindent\begin{tikzpicture}[remember picture,overlay]
\draw[white,thick]  (2.032,8.965) rectangle (3.18,8.2) node[below=12.0 ,pos=.55]{1B};
\draw[white,thick]  (3.87,7.775) rectangle (7.26,5.58) node[below=40.0 ,pos=.5]{1A};
\end{tikzpicture}%
\vspace{-0.5cm}
\caption{Overall clustering result with $i_{\mathrm{min}} = 0.29$, $M = 25\,519$ and the features in the rows below. The feature rows use the same color scheme as introduced with $\bm{P}_\mathrm{o,1}$, where blue equals the lowest, yellow the highest value. Still, the color is set according the min and max value of the respective feature. The bottom row depicts the scenery type.}
\label{fig:fig_results1_1}
\vspace{-0.5cm}
\end{figure}

In Fig. \ref{fig:fig_results1_1} the feature values from the diagonal datapoints are introduced. Their values are depicted with the same colormap scheme as the $\bm{P}_{\mathrm{o}}$ matrix. However, the range of colors is calibrated according to the quantitative value of the respective feature. The four velocity features share a common range for the quantitative minimum and maximum values. All other feature rows are set independently according to their quantitative values. Finally, the last row depicts the scenery type using the color coding described above. This row shall support readers' understanding of the clustering results. It has to be stressed that the clustering in the upper figure has been performed without any knowledge about the scenarios in an unsupervised automatical way. 
%Table no. \ref{table:table1} depicts the rounded boundary values for each feature. Velocity related features are given in SI units and angles in degrees.
%
%\begin{tabular}{|c|c|c|c|c|c|c|}
%
%\hline 
%• & $v$ & $b$ & $\delta_{rel}$ & $r$ & $v_{lim}$ & $n_{L}$ \\  
%\hline 
%min & 0 & 0 & 0 & 10 & 8 & 1 \\ 
%\hline 
%max & 60 & 1 & 180 & 11111 & 80 & 4 \\ 
%\hline
%\label{table:table1}
%\end{tabular} 
  
The speed limit ($v_\mathrm{lim}$) and the number of lanes ($n_\mathrm{L}$) are usually higher on highways than for inner city scenarios as depicted in Fig. \ref{fig:fig_results1_1}. Note, that one highway segment consists of only two lanes and a reduced speed limit, which represents a construction site where the third lane is closed. The average velocities are considerable higher on highways and contribute to the separation during the clustering process. Additionally, the relative angles ($\delta_\mathrm{rel}$) are usually in a range of \unit[0]{$^{\circ}$} to \unit[30]{$^{\circ}$}, as on highways rear-end accidents or accidents due to lane change maneuvers occur. One example of a separated cluster due to the object features is marked in box no. 1B. On the left handed side of box no. 1B both vehicles drive with high velocities. On the right handed side the target vehicle drives at a clearly lower velocity.

\begin{figure}
\par\medskip
\begin{scriptsize}
	\parbox[c]{0.10\columnwidth}{\raggedright $\bm{P}_\mathrm{o,2}$}%
	\fbox{\parbox[c]{0.89\columnwidth}{\raggedleft \def\svgwidth{0.89\columnwidth} 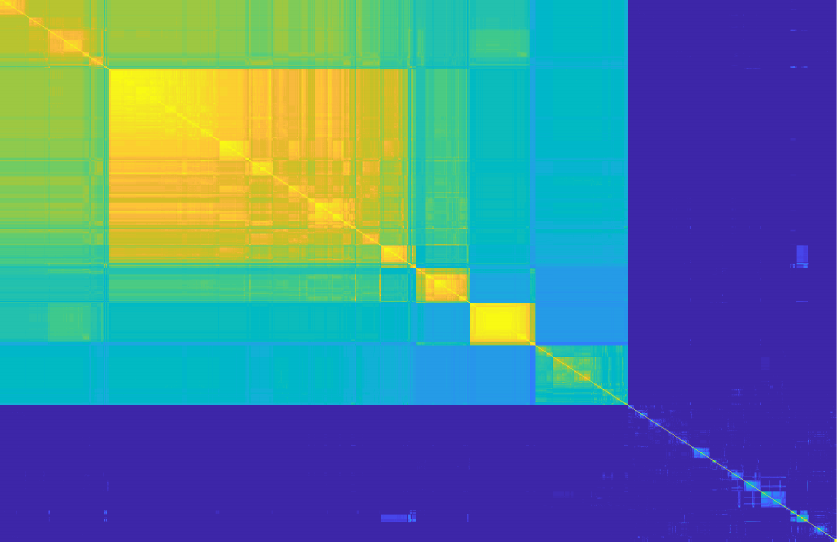}}\\[1ex]
	\parbox[c]{0.10\columnwidth}{\raggedright $v_{\mathrm{eg}_\mathrm{t-2}}$}%
	\fbox{\parbox[c]{0.89\columnwidth}{\raggedleft \def\svgwidth{0.89\columnwidth} 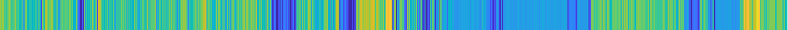}}\\[0.5ex]
	\parbox[c]{0.10\columnwidth}{\raggedright $v_{\mathrm{eg}_\mathrm{t0}}$}%
	\fbox{\parbox[c]{0.89\columnwidth}{\raggedleft \def\svgwidth{0.89\columnwidth} 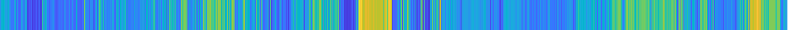}}\\[0.5ex]
	\parbox[c]{0.10\columnwidth}{\raggedright $b_{\mathrm{eg}}$}%
	\fbox{\parbox[c]{0.89\columnwidth}{\raggedleft \def\svgwidth{0.89\columnwidth} 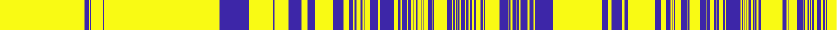}}\\[0.5ex]
	\parbox[c]{0.10\columnwidth}{\raggedright $v_{\mathrm{tg}_\mathrm{t-2}}$}%
	\fbox{\parbox[c]{0.89\columnwidth}{\raggedleft \def\svgwidth{0.89\columnwidth} 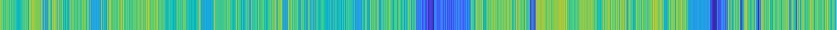}}\\[0.5ex]
	\parbox[c]{0.10\columnwidth}{\raggedright $v_{\mathrm{tg}_\mathrm{t0}}$}%
	\fbox{\parbox[c]{0.89\columnwidth}{\raggedleft \def\svgwidth{0.89\columnwidth} 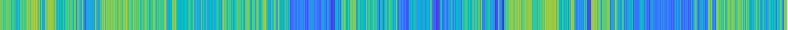}}\\[0.5ex]
	\parbox[c]{0.10\columnwidth}{\raggedright $b_{\mathrm{tg}}$}%
	\fbox{\parbox[c]{0.89\columnwidth}{\raggedleft \def\svgwidth{0.89\columnwidth} 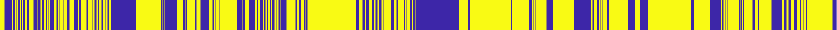}}\\[0.5ex]
	\parbox[c]{0.10\columnwidth}{\raggedright $\delta_\mathrm{rel}$}%
	\fbox{\parbox[c]{0.89\columnwidth}{\raggedleft \def\svgwidth{0.89\columnwidth} 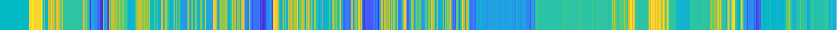}}\\[0.5ex]
	\parbox[c]{0.10\columnwidth}{\raggedright $r$}%
	\fbox{\parbox[c]{0.89\columnwidth}{\raggedleft \def\svgwidth{0.89\columnwidth} 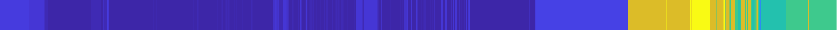}}\\[0.5ex]
	\parbox[c]{0.10\columnwidth}{\raggedright $v_\mathrm{lim}$}%
	\fbox{\parbox[c]{0.89\columnwidth}{\raggedleft \def\svgwidth{0.89\columnwidth} 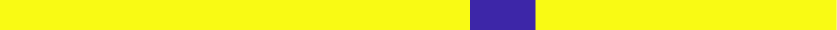}}\\[0.5ex]
	\parbox[c]{0.10\columnwidth}{\raggedright $n_\mathrm{L}$}%
	\fbox{\parbox[c]{0.89\columnwidth}{\raggedleft \def\svgwidth{0.89\columnwidth} 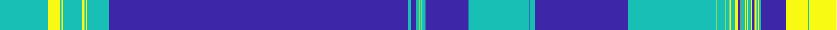}}\\[1.2ex]
	\parbox[c]{0.10\columnwidth}{\raggedright type}%
	\fbox{\parbox[c]{0.89\columnwidth}{\raggedleft \def\svgwidth{0.89\columnwidth} 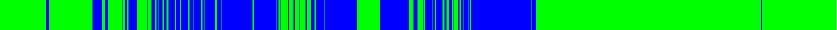}}\\[0.5ex]
	
\end{scriptsize}
\noindent\begin{tikzpicture}[remember picture,overlay]
\draw[white,thick]  (0.888,9.73) rectangle (6.63,6.01) node[below=53.0 ,pos=.5]{2};
\end{tikzpicture}%
\vspace{-0.5cm}
\caption{The resulting proximity matrix $\bm{P}_\mathrm{o,2}$ with $M = 9\,876$ which is generated as a second iteration step by re-clustering the data subset of box no. 1. The impurity threshold is set to $i_{\mathrm{min}} = 0.29$. Box no. 2 will be once more re-clustered for a final iteration step. Note that the color range of the features values are adapted to this data subset.}
\label{fig:fig_results2}
\vspace{-0.5cm}
\end{figure}
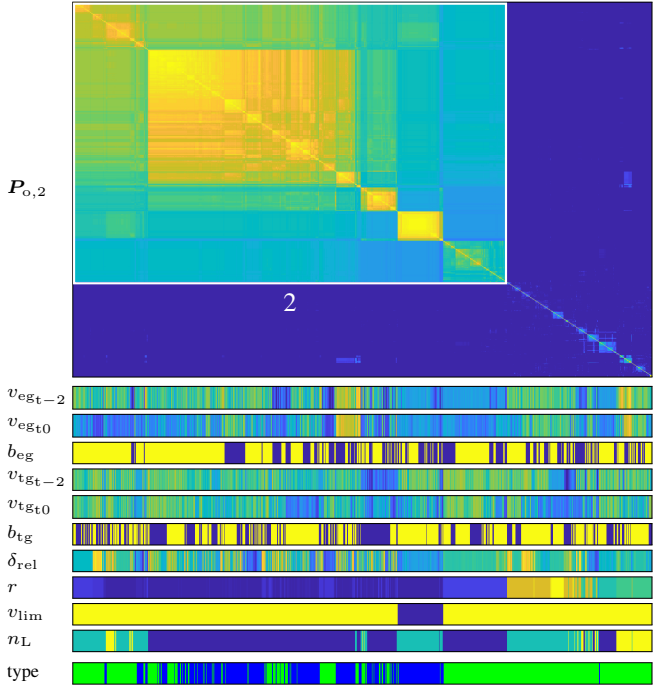

According to the set of features, their quantitative values differ remarkably between the highway and inner city scenarios. Therefore, another clustering iteration for the sub data set located in box no. 1A in Fig. \ref{fig:fig_results1_1} is performed. The datapoints are exclusively inner city scenarios. The parameter $i_{\mathrm{min}}$ can be set independently to the prior clustering. By re-running the clustering the strength of the data adaptiveness of the URF algorithm is utilized. The new clusters in Fig. \ref{fig:fig_results2} emerge due to the actual quantitative range of the feature values. The re-clustering can be performed for all relevant clusters. The discussion is continued on the base of the new matrix $\bm{P}_\mathrm{o,2}$ which is depicted in Fig. \ref{fig:fig_results2}. Note that the matrix $\bm{P}_\mathrm{o,2}$ is derived from the data set in box. no 1A.

In Fig. \ref{fig:fig_results2} one major cluster in box no. 2 can be detected. All the other pixels have a low similarity for $i_{\mathrm{min}} = 0.29$. For those pixels a change of $i_{\mathrm{min}}$ or an additional clustering process might reveal some common structure. If this is not the case, they would be of high interest, since their individual structure would be quite unique and represent rarely appearing traffic scenarios. As pronounced in \cite{DBLP:journals/corr/ZhaoP17aa}, the number of such scenarios should be extended by varying the features in order to validate the vehicles' behavior. As a result, the validation process can be accelerated.  

\begin{figure}
\par\medskip
\begin{scriptsize}
	\parbox[c]{0.10\columnwidth}{\raggedright $\bm{P}_\mathrm{o,3}$}%
	\fbox{\parbox[c]{0.89\columnwidth}{\raggedleft \def\svgwidth{0.89\columnwidth} 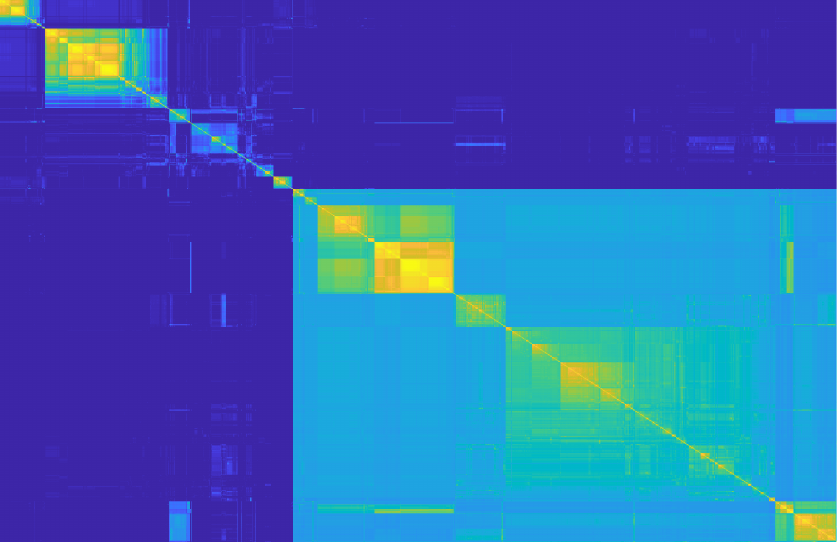}}\\[1ex]
	\parbox[c]{0.10\columnwidth}{\raggedright $v_{\mathrm{eg}_\mathrm{t-2}}$}%
	\fbox{\parbox[c]{0.89\columnwidth}{\raggedleft \def\svgwidth{0.89\columnwidth} 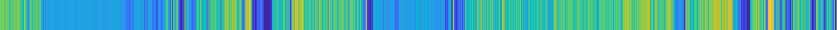}}\\[0.5ex]
	\parbox[c]{0.10\columnwidth}{\raggedright $v_{\mathrm{eg}_\mathrm{t0}}$}%
	\fbox{\parbox[c]{0.89\columnwidth}{\raggedleft \def\svgwidth{0.89\columnwidth} 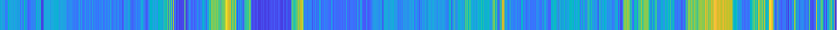}}\\[0.5ex]
	\parbox[c]{0.10\columnwidth}{\raggedright $b_{\mathrm{eg}}$}%
	\fbox{\parbox[c]{0.89\columnwidth}{\raggedleft \def\svgwidth{0.89\columnwidth} 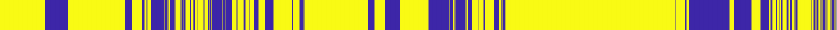}}\\[0.5ex]
	\parbox[c]{0.10\columnwidth}{\raggedright $v_{\mathrm{tg}_\mathrm{t-2}}$}%
	\fbox{\parbox[c]{0.89\columnwidth}{\raggedleft \def\svgwidth{0.89\columnwidth} 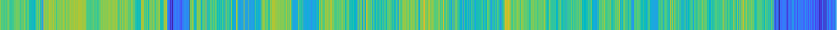}}\\[0.5ex]
	\parbox[c]{0.10\columnwidth}{\raggedright $v_{\mathrm{tg}_\mathrm{t0}}$}%
	\fbox{\parbox[c]{0.89\columnwidth}{\raggedleft \def\svgwidth{0.89\columnwidth} 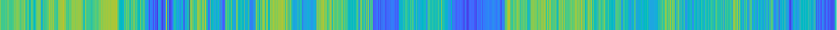}}\\[0.5ex]
	\parbox[c]{0.10\columnwidth}{\raggedright $b_{\mathrm{tg}}$}%
	\fbox{\parbox[c]{0.89\columnwidth}{\raggedleft \def\svgwidth{0.89\columnwidth} 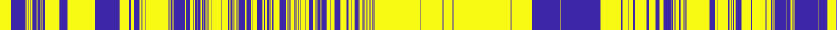}}\\[0.5ex]
	\parbox[c]{0.10\columnwidth}{\raggedright $\delta_\mathrm{rel}$}%
	\fbox{\parbox[c]{0.89\columnwidth}{\raggedleft \def\svgwidth{0.89\columnwidth} 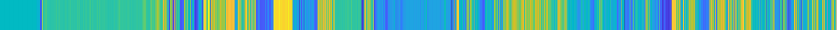}}\\[0.5ex]
	\parbox[c]{0.10\columnwidth}{\raggedright $r$}%
	\fbox{\parbox[c]{0.89\columnwidth}{\raggedleft \def\svgwidth{0.89\columnwidth} 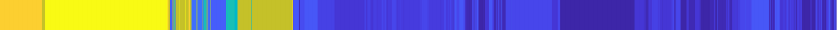}}\\[0.5ex]
	\parbox[c]{0.10\columnwidth}{\raggedright $v_\mathrm{lim}$}%
	\fbox{\parbox[c]{0.89\columnwidth}{\raggedleft \def\svgwidth{0.89\columnwidth} 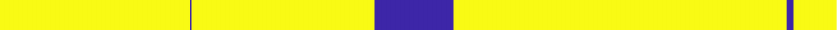}}\\[0.5ex]
	\parbox[c]{0.10\columnwidth}{\raggedright $n_\mathrm{L}$}%
	\fbox{\parbox[c]{0.89\columnwidth}{\raggedleft \def\svgwidth{0.89\columnwidth} 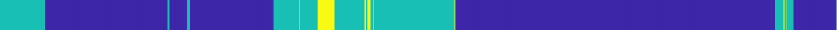}}\\[1.2ex]
	\parbox[c]{0.10\columnwidth}{\raggedright type}%
	\fbox{\parbox[c]{0.89\columnwidth}{\raggedleft \def\svgwidth{0.89\columnwidth} 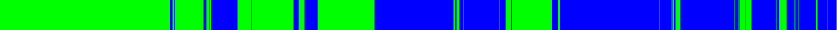}}\\[0.5ex]
	
\end{scriptsize}
\noindent\begin{tikzpicture}[remember picture,overlay]
\draw[red,thick] (0.86,0.671) -- (0.86,9.7078);
\draw[red,thick] (1.106,0.671) -- (1.106,9.7078);
\draw[white,thick]  (0.86,9.7078) rectangle (1.106,9.54) node[right=1.5,pos=.8]{3A};
\draw[red,thick] (3.928,0.671) -- (3.928,7.7303);
\draw[red,thick] (4.205,0.671) -- (4.205,7.7303);
\draw[white,thick]  (3.928,7.7303) rectangle (4.205,7.5393) node[left=2.5,pos=.5]{3B};
\draw[red,thick] (4.55,0.671) -- (4.55,7.3);
\draw[red,thick] (4.8,0.671) -- (4.8,7.3);
\draw[white,thick]  (4.55,7.33) rectangle (4.8,7.16) node[right=7.5,pos=.5]{3C};
\end{tikzpicture}
\vspace{-0.5cm}
\caption{The resulting distance matrix $\bm{P}_\mathrm{o,3}$ with $i_{\mathrm{min}} = 0.34$ and $M = 7\,427$. This distance matrix results from re-clustering the datapoints of box no. 2.}
\label{fig:fig_results4}
%\vspace{-0.3cm}
\end{figure}
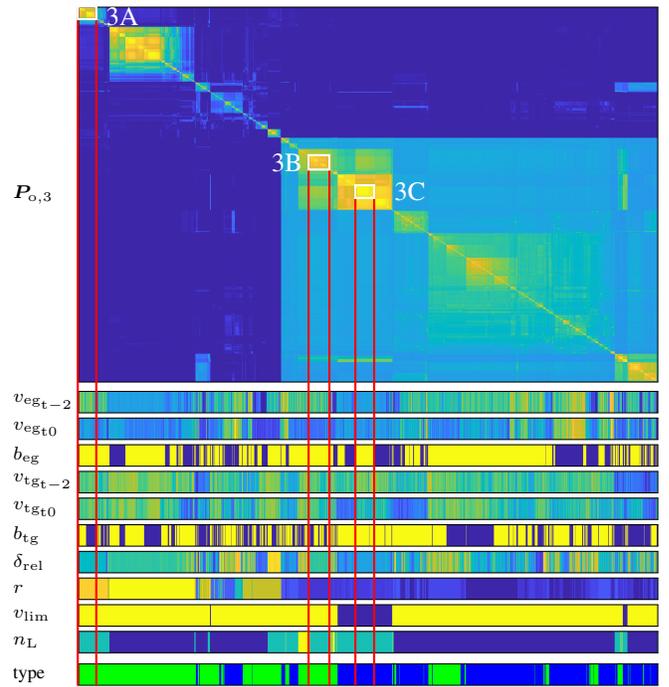

\begin{figure}
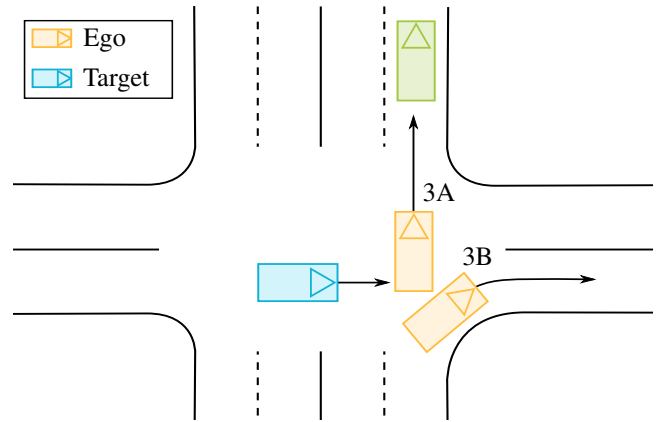
\caption{Visualization of the scenarios in box no. 3A and 3B.}
\label{fig:crossing}
\vspace{-0.5cm}
\end{figure}

For this paper, the workflow of re-clustering the scenarios with high similarities is continued. The datapoints of box no. 2 are chosen. The resulting proximity matrix $\bm{P}_\mathrm{o,3}$ is depicted in Fig. \ref{fig:fig_results4}. The discussion follows based on the boxes in Fig. \ref{fig:fig_results4}. Boxes 3A and 3B contain crossing scenarios, the box no. 3C roundabout scenarios. In the scenarios of 3A and 3B the target vehicle drives considerable faster than the ego vehicle, which starts a braking maneuver. The main difference can be explained with the road segment radius. In box no. 3A the ego vehicle passes the crossing on a rather straight inner city road with a radius around 130\,m. In box no. 3B the ego vehicle performs a turn into another road, indicated by a road segment radius of around 15\,m. In both cases the vehicle brakes, due to the turning maneuver (3B) or because of objects ahead (3A). The relative angle is around \unit[90]{$^{\circ}$} in 3A and \unit[110]{$^{\circ}$} in 3B. In both groups the scenarios end in a side crash. A sketch of the scenery is depicted in Fig. \ref{fig:crossing}. Now box no. 3C is taken into consideration. The speed limit at these roundabout scenarios is lower. All other feature values are comparable to 3A and 3B. The roundabout scenarios in 3C are side crashes where one of the vehicles enters the roundabout. 
If one thinks about creating test cases, which is one of the important applications of the presented methodology, those three clusters can be combined. Performing the tests on a crossing scenery and varying the vehicle trajectories, one is able to gather data for all three clusters. This is a simplified example, how the designing of test cases can benefit from the clustering process.

Summarizing this section, the proposed method is fed with a data set $\mathcal{D}$. No knowledge about the traffic scenarios is assumed by the proposed method. Similarities are revealed automatically and the clusters can be used to generate templates. In addition to that, inter-cluster similarities can be detected. The clustering method allows different insights by varying $i_{\mathrm{min}}$. In the case of a highly heterogen data set $\mathcal{D}$, but inherent structures within a cluster, an iterated clustering process reveals deeper insights as shown in this section. The set of ten features and the eight simulated sceneries is sufficient to demonstrate the potential of the clustering method. Nevertheless, more work and research needs to be invested to cluster complex real world applications. 

\section{CONCLUSIONS}
\label{conclusions_section}
This work describes an automatic method to find clusters in traffic situations using a data adaptive similarity measure that is provided by a modified URF algorithm. The modified URF followed by hierarchical clustering is capable of providing an adjustable insight into data inherent structures. According to the balanced noising, the input space size can be very high. The metaparamter $i_{\mathrm{min}}$ can be used in order to set the granularity of the structured data. The generated scenario clusters can be used as labels for a following classification, i.\,e. supervised learning, task. This way, the effort for labeling is greatly reduced. This approach yields to a semi-supervised learning.

One main drawback of the proposed method is the quadratic increase of the proximity matrix depending on the data set size. Thus, working with a reduced data size in combination with the resulting semi-supervised learning technique, is part of the current research on the topic.

The clusters can be utilized to determine templates of representative traffic situations. 
Since many years the template approach is widely used in the passive safety validation process. Approvals or ratings are often based on a set of test cases. The NCAP crash tests are a well known example for that. In order to achieve a rating, a defined set of representative crash scenarios is examined. Test cases are adjusted with increasing experience and technological progress.

The traffic scenarios templates can be deployed in different manners for the validation process. For example, templates with only few datapoints represent rare situations on the road and are therefore valuable. Their specific setup can be analyzed and used in the further development process. By varying the features in small steps, e.g. on a test track or on a simulation level, a deeper knowledge of the vehicle behavior can be gained. Thus, they contribute to the testing coverage. The effort is strongly reduced compared to a classical kilometer-based validation approach. Alternatively, combining the feature set with the corresponding GPS position, one can check on which roads certain scenarios took place. Templates with only few datapoints are then favored in a route planning algorithm. An optimized field test cycle based on the cluster coverage can be achieved.

As traffic scenarios can be generated in a simulation environment with many degrees of freedom, clusters can be found which are unlikely, but yet not impossible to appear in the real world. Obviously, those scenarios should also be tested with the vehicle to enhance the test coverage. 
Once a pool of templates is available, future vehicle models can benefit, as those templates are valuable for all development phases, from software-in-the-loop tests to the final phase of field tests on the roads. That way, the necessary amount for driving can be shrinked. It should be mentioned, that the learning process itself shall be continuous, because it is likely to find new templates with a rising range of driven or simulated kilometers. Scenarios which are not assigned to any cluster form a new template. Consequently, a metric for thresholding similarity measures has to be developed.

Thinking of other research topics which can benefit from the method, the crash severity prediction example is reconsidered. The templates resulting from the clustering process represent different classes of crash scenarios. Once the classes are known from the training process, predictive algorithms assign a scenario to the corresponding class just before the collision occurs. The classification task can be fulfilled computational fast because the input data is distilled to a feature vector. Actions of safety systems can be adjusted with that knowledge. For example, the inflation pressure of airbags can be varied according to the predicted crash severity. Generally speaking, applications with multi-variable data, which has an inherent structure can be clustered in a data driven way by using the proposed method.

\addtolength{\textheight}{-12cm}   % This command serves to balance the column lengths

%%%%%%%%%%%%%%%%%%%%%%%%%%%%%%%%%%%%%%%%%%%%%%%%%%%%%%%%%%%%%%%%%%%%%%%%%%%%%%%%

%%%%%%%%%%%%%%%%%%%%%%%%%%%%%%%%%%%%%%%%%%%%%%%%%%%%%%%%%%%%%%%%%%%%%%%%%%%%%%%%

%%%%%%%%%%%%%%%%%%%%%%%%%%%%%%%%%%%%%%%%%%%%%%%%%%%%%%%%%%%%%%%%%%%%%%%%%%%%%%%%

\section*{ACKNOWLEDGMENT}
The authors acknowledge the financial support by the Federal Ministry of Education and Research of
Germany (BMBF) in the framework of FH-Impuls (project number 03FH7I02IA). The authors thank Dr. Antesberger and Dr. Schlicht from the AUDI AG department for Testing Total Vehicle for supporting this work.

\bibliographystyle{IEEEtran}
\bibliography{ref}
\end{document}